\documentclass[twocolumn,11pt]{article}
\usepackage{times}
\usepackage{graphicx}
\begin{document}
\global\def\refname{{\normalsize \it References:}}
\baselineskip 12.5pt
%
%
%
\title{\LARGE \bf Equilibrium Distributions in Open and Closed Statistical Systems}

\date{}

\author{\hspace*{-10pt}
\begin{minipage}[t]{2.0in} \footnotesize \baselineskip 10.5pt
\centerline{\normalsize RICARDO L\'OPEZ-RUIZ}
\centerline{Universidad de Zaragoza}
\centerline{Department of Computer Science and BIFI}
\centerline{Campus San Francisco, E-50009 Zaragoza}
\centerline{SPAIN}
\centerline{rilopez@unizar.es}
\end{minipage} 
\begin{minipage}[t]{2.0in} \footnotesize \baselineskip 10.5pt
\centerline{\normalsize JAIME SA\~NUDO}
\centerline{Universidad de Extremadura}
\centerline{Department of Physics and BIFI}
\centerline{Avda. de Elvas, E-06071 Badajoz}
\centerline{SPAIN}
\centerline{jsr@unex.es}
\end{minipage}
\begin{minipage}[t]{2.0in} \footnotesize \baselineskip 10.5pt
\centerline{\normalsize XAVIER CALBET}
\centerline{Instituto de Biocomputaci\'on y F\'isica de}
\centerline{Sistemas Complejos (BIFI), Edificio Cervantes}
\centerline{Corona de Arag\'on 42, E-50009 Zaragoza}
\centerline{SPAIN}
\centerline{xcalbet@googlemail.com}
\end{minipage}
%
%
\\ \\ \hspace*{-10pt}
\begin{minipage}[b]{6.9in} \normalsize
\baselineskip 12.5pt {\it Abstract:}
{\small In this communication, the derivation of the Boltzmann-Gibbs and the Maxwellian distributions
is presented from a geometrical point of view under the hypothesis of equiprobability. 
It is shown that both distributions can be obtained by working out the properties 
of the volume or the surface of the respective geometries delimited in phase space
by an additive constraint. That is, the asymptotic equilibrium distributions in the thermodynamic limit 
are independent of considering open or closed homogeneous statistical systems.} 
\\ [4mm] {\it Key--Words:}
Equiprobability, asymptotic equilibrium distributions, geometrical derivation
\end{minipage}
\vspace{-10pt}}

\maketitle

\thispagestyle{empty} \pagestyle{empty}
%
%

\section{Introduction}
\label{S1} \vspace{-4pt}

In this paper, different classical results \cite{huang1987,yako2009} are obtained from a 
geometrical interpretation of different multi-agent systems evolving in 
phase space under the hypothesis of equiprobability. 

We start by deriving in section \ref{S2} the Boltzmann-Gibbs (exponential) distribution 
by means of the geometrical properties of the volume of an $N$-dimensional pyramid.
The same result is obtained when the calculation is performed over the surface of 
a such $N$-dimensional body. In both cases, the motivation is a multi-agent economic system
with an open or closed economy, respectively.  

Also, the Maxwellian (Gaussian) distribution is derived in section \ref{S3}
from geometrical arguments over the volume or the surface of an $N$-sphere.
Here, the motivation is a multi-particle gas system in contact
with a heat reservoir (non-isolated or open system) or with a fixed energy
(isolated or closed system), respectively. 

Last section contains our conclusions.

\section{Derivation of the Boltzmann-Gibbs Distribution}
\label{S2} \vspace{-4pt}

\subsection{Multi-agent economic open systems}

Here we assume $N$ agents, each one with coordinate $x_i$, $i=1,\ldots,N$, 
with $x_i\geq 0$ representing the wealth or money of the agent $i$,
and a total available amount of money $E$:
\begin{equation}
x_1+x_2+\cdots +x_{N-1}+x_N \leq E.
\label{eq-e}
\end{equation} 
Under random or deterministic evolution rules for the exchanging of money among agents,
let us suppose that this system evolves in the interior of the $N$-dimensional pyramid 
given by Eq. (\ref{eq-e}). The role of a heat reservoir, that in this model
supplies money instead of energy, could be played by the state or by the bank system 
in western societies.  
The formula for the volume $V_N(E)$ of an equilateral $N$-dimensional pyramid 
formed by $N+1$ vertices linked by $N$ perpendicular sides of length $E$ is
\begin{equation}
V_N(E) = {E^N\over N!}.
\label{eq-S_n1}
\end{equation}
We suppose that each point on the $N$-dimensional pyramid is equiprobable, 
then the probability $f(x_i)dx_i$ of finding 
the agent $i$ with money $x_i$ is proportional to the 
volume formed by all the points into the $(N-1)$-dimensional pyramid 
having the $i$th-coordinate equal to $x_i$. 
We show now that $f(x_i)$ is the Boltzmann factor
(or the Maxwell-Boltzmann distribution), with the normalization condition
\begin{equation}
\int_{0}^Ef(x_i)dx_i = 1.
\label{eq-p_n1}
\end{equation}

If the $i$th agent has coordinate $x_i$, the $N-1$ remaining agents 
share, at most, the money $E-x_i$ on the $(N-1)$-dimensional pyramid
\begin{equation}
x_1+x_2 \cdots +x_{i-1} + x_{i+1} \cdots +x_N\leq E-x_i,
\label{eq-e1}
\end{equation} 
whose volume is $V_{N-1}(E-x_i)$. 
It can be easily proved that
\begin{equation}
V_N(E) = \!\int_{0}^{E}\!V_{N-1}(E-x_i) {dx_i }.
\label{eq-theta11}
\end{equation}

Hence, the volume of the $N$-dimensional pyramid for which the $i$th 
coordinate is between $x_i$ and $x_i+dx_i$ is $V_{N-1}(E-x_i)dx_i$.
We normalize it to satisfy Eq.~(\ref{eq-p_n1}), and obtain
\begin{equation}
f(x_i) = {V_{N-1}(E-x_i)\over V_N(E)},
\label{eq-f_n1}
\end{equation}
whose final form, after some calculation is
\begin{equation}
f(x_i) = NE^{-1}\Big(1-{x_i\over E} \Big)^{N-1},
\label{eq-mm1}
\end{equation}
If we call $\epsilon$ the mean wealth per agent, 
$E=N\epsilon$, then in the limit of large $N$ 
we have
\begin{equation}
\lim_{N\gg 1}\left(1-{x_i\over E}\right)^{N-1}
\simeq e^{-{x_i/\epsilon}}.
\label{eq-ee1}
\end{equation}
The Boltzmann factor $e^{-{x_i/\epsilon}}$ is found 
when $N\gg 1$ but, even for small $N$, it can be a good approximation 
for agents with low wealth. After substituting Eq.~(\ref{eq-ee1})
into Eq.~(\ref{eq-mm1}), we obtain the Maxwell-Boltzmann distribution 
in the asymptotic regime $N\rightarrow\infty$ (which also implies $E\rightarrow\infty$):
\begin{equation}
f(x)dx = {1\over \epsilon}\,e^{-{x/\epsilon}}dx,
\label{eq-gauss11}
\end{equation}
where the index $i$ has been removed because the distribution is the same for each agent, 
and thus the wealth distribution can be obtained by averaging over all the agents. 
This distribution has been found to fit the real distribution of incomes 
in western societies \cite{yako2001}.

This means that the geometrical image of the volume-based statistical ensemble
allows us to recover the same result than that obtained 
from the microcanonical ensemble \cite{lopez2008} that we show in the next section.

\subsection{Multi-agent economic closed systems}

Here, we derive the Boltzmann-Gibbs distribution by considering the system 
in isolation, that is, a closed economy. Without loss of generality, 
let us assume $N$ interacting economic agents, each one with coordinate 
$x_i$, $i=1,\ldots,N$, with $x_i\geq 0$, and where $x_i$ represents an amount of money. 
If we suppose now that the total amount of money $E$ is conserved,
\begin{equation}
x_1+x_2+\cdots +x_{N-1}+x_N = E,
\label{eq-E}
\end{equation} 
then this isolated system 
evolves on the positive part of an equilateral $N$-hyperplane. 
The surface area $S_N(E)$ of an equilateral 
$N$-hyperplane of side $E$ is given by
\begin{equation}
S_N(E) = {\sqrt{N}\over (N-1)!}\;E^{N-1}.
\label{eq-S_n2}
\end{equation}
Different rules, deterministic or random, for the exchange of money between agents 
can be given \cite{yako2009}. Depending on these rules, the system can visit the 
$N$-hyperplane in an equiprobable manner or not. 
If the ergodic hypothesis is 
assumed, each point on the $N$-hyperplane is equiprobable. 
Then the probability $f(x_i)dx_i$ of finding 
agent $i$ with money $x_i$ is proportional to the 
surface area formed by all the points on the $N$-hyperplane having the $i$th-coordinate 
equal to $x_i$. We show that $f(x_i)$ is the Boltzmann factor (Boltzmann-Gibbs distribution),
with the normalization condition (\ref{eq-p_n1}).

If the $i$th agent has coordinate $x_i$, the $N-1$ remaining agents 
share the money $E-x_i$ on the $(N-1)$-hyperplane
\begin{equation}
x_1+x_2 \cdots +x_{i-1} + x_{i+1} \cdots +x_N= E-x_i,
\label{eq-e11}
\end{equation} 
whose surface area is $S_{N-1}(E-x_i)$. 
If we define the coordinate $\theta_N$ as satisfying
\begin{equation}
\sin\theta_N = \sqrt{N-1 \over N},
\label{eq-theta}
\end{equation}
it can be easily shown that
\begin{equation}
S_N(E) = \!\int_{0}^{E}\!S_{N-1}(E-x_i) {dx_i \over \sin\theta_N}.
\label{eq-theta122}
\end{equation}

Hence, the surface area of the $N$-hyperplane for which the $i$th coordinate is
between $x_i$ and $x_i+dx_i$ is proportional to $S_{N-1}(E-x_i)dx_i/\sin\theta_N$.
If we take into account the normalization condition (\ref{eq-p_n1}), we obtain
\begin{equation}
f(x_i) = {1\over S_N(E)}
{S_{N-1}(E-x_i)\over \sin\theta_N},
\label{eq-f_n}
\end{equation}
whose form after some calculation is
\begin{equation}
f(x_i) = (N-1)E^{-1}\Big(1-{x_i\over E} \Big)^{N-2},
\label{eq-mmm}
\end{equation}
If we call $\epsilon$ the mean wealth per agent, 
$E=N\epsilon$, then in the limit of large $N$ 
we have
\begin{equation}
\lim_{N\gg 1}\left(1-{x_i\over E}\right)^{N-2}
\simeq e^{-{x_i/\epsilon}}.
\label{eq-eee}
\end{equation}
As in the former section, the Boltzmann factor $e^{-{x_i/\epsilon}}$ is found 
when $N\gg 1$ but, even for small $N$, it can be a good approximation 
for agents with low wealth. After substituting Eq.~(\ref{eq-eee})
into Eq.~(\ref{eq-mmm}), we obtain the Boltzmann distribution (\ref{eq-gauss11})
in the limit $N\rightarrow\infty$ (which also implies $E\rightarrow\infty$).
This asymptotic result reproduces the distribution of real economic data \cite{yako2001}
and also the results obtained in several models of economic agents
with random exchange interactions \cite{yako2009}. 

Depending on the physical situation, the mean wealth per agent $\epsilon$
takes different expressions and interpretations. 
For instance, 
we can calculate the dependence of $\epsilon$ on the temperature, which
in the microcanonical ensemble is defined by the derivative of the entropy
with respect to the energy. The entropy can be written as $S=-kN\int_{0} 
^{\infty} f(x)\ln f(x)\,dx$, where $f(x)$ is given by Eq.~(\ref{eq-gauss11})
and $k$ is Boltzmann's constant. 
If we recall that $\epsilon=E/N$, we obtain
\begin{equation}
S(E)= kN\ln {E\over N} + kN.
\end{equation}
The calculation of the temperature $T$ gives
\begin{equation}
T^{-1}= \left({\partial S\over \partial E} \right)_N = {kN\over E} = {k\over \epsilon}.
\end{equation}
Thus $\epsilon=kT$, and the Boltzmann distribution 
is obtained in its usual form:
\begin{equation}
f(x)dx = {1\over kT}\,e^{-x/kT}dx.
\end{equation}

\section{Derivation of the Maxwellian Distribution}
\label{S3} \vspace{-4pt}

\subsection{Multi-particle open systems}

Let us suppose a one-dimensional ideal gas of $N$ non-identical 
classical particles with masses $m_i$, with $i=1,\ldots,N$, and total 
maximum energy $E$. If particle
$i$ has a momentum $m_iv_i$, we define a kinetic energy:
\begin{equation}
K_i \equiv p_i^2 \equiv {1 \over 2}{ m_iv_i^2},
\label{eq-p_i}
\end{equation} 
where $p_i$ is the square root of the kinetic energy $K_i$. 
If the total maximum energy is defined as $E \equiv R^2$, we have 
\begin{equation}
p_1^2+p_2^2+\cdots +p_{N-1}^2+p_N^2 \leq R^2.
\label{eq-Ee}
\end{equation} 
We see that the system has accessible states with different energy, which can be 
supplied by a heat reservoir. These states are all those enclosed into the volume 
of the $N$-sphere given by Eq. (\ref{eq-Ee}). 
The formula for the volume $V_N(R)$
of an $N$-sphere of radius $R$ is
\begin{equation}
V_N(R) = {\pi^{N\over 2}\over \Gamma({N\over 2}+1)}R^{N},
\label{eq-S_n3}
\end{equation}
where $\Gamma(\cdot)$ is the gamma function. If we suppose that each point
into the $N$-sphere is equiprobable, then the probability $f(p_i)dp_i$ of finding 
the particle $i$ with coordinate $p_i$ (energy $p_i^2$) is proportional to the 
volume formed by all the points on the $N$-sphere having the $i$th-coordinate 
equal to $p_i$. 
We proceed to show that $f(p_i)$ is the Maxwellian 
distribution, with the normalization condition
\begin{equation}
\int_{-R}^Rf(p_i)dp_i = 1.
\label{eq-p_n2}
\end{equation}

If the $i$th particle has coordinate $p_i$, the $(N-1)$ remaining particles 
share an energy less than the maximum energy $R^2-p_i^2$ on the $(N-1)$-sphere
\begin{equation}
p_1^2+p_2^2 \cdots +p_{i-1}^2 + p_{i+1}^2 \cdots +p_N^2 \leq R^2-p_i^2,
\label{eq-E12}
\end{equation} 
whose volume is $V_{N-1}(\sqrt{R^2-p_i^2})$. 
It can be easily proved that
\begin{equation}
V_N(R) = \!\int_{-R}^{R}\!V_{N-1}(\sqrt{R^2-p_i^2})dp_i.
\label{eq-theta13}
\end{equation}
Hence, the volume of the $N$-sphere for which the $i$th coordinate is
between $p_i$ and $p_i+dp_i$ is $V_{N-1}(\sqrt{R^2-p_i^2})dp_i$.
We normalize it to satisfy Eq.~(\ref{eq-p_n2}), and obtain
\begin{equation}
f(p_i) = {V_{N-1}(\sqrt{R^2-p_i^2})\over V_N(R)},
\label{eq-f_n2}
\end{equation}
whose final form, after some calculation is
\begin{equation}
f(p_i) = C_N R^{-1}\Big(1-{p_i^2\over R^2} \Big)^{N-1\over 2},
\label{eq-mmm1}
\end{equation}
with
\begin{equation}
C_N = {1\over\sqrt{\pi}}{\Gamma({N+2\over 2})\over \Gamma({N+1\over 2})}.
\label{eq-cnn}
\end{equation}
For $N\gg 1$, Stirling's approximation can be applied to 
Eq.~(\ref{eq-cnn}), leading to
\begin{equation}
\lim_{N\gg 1} C_N \simeq {1\over\sqrt{\pi}}\sqrt{N\over 2}.
\label{eq-ccc}
\end{equation}
If we call $\epsilon$ the mean energy per particle, 
$E=R^2=N\epsilon$, then in the limit of large $N$ we have
\begin{equation}
\lim_{N\gg 1}\left(1-{p_i^2\over R^2}\right)^{N-1\over 2}
\simeq e^{-{p_i^2/2\epsilon}}.
\label{eq-eee1}
\end{equation}
The factor $e^{-{p_i^2/2\epsilon}}$ is found 
when $N\gg 1$ but, even for small $N$, it can be a good approximation 
for particles with low energies.
After substituting Eqs.~(\ref{eq-ccc})--(\ref{eq-eee1})
into Eq.~(\ref{eq-mmm1}), we obtain the Maxwellian distribution in the asymptotic regime $N\rightarrow\infty$ 
(which also implies $E\rightarrow\infty$):
\begin{equation}
f(p)dp = \sqrt{1\over 2\pi\epsilon}\,e^{-{p^2/2\epsilon}}dp,
\label{eq-gauss}
\end{equation}
where the index $i$ has been removed because the distribution is the same for each particle, 
and thus the velocity distribution can be obtained by averaging
over all the particles. 

This newly shows that the geometrical image of the volume-based statistical ensemble
allows us to recover the same result than that obtained 
from the microcanonical ensemble \cite{lopez2007} that it is presented in the next section.

\subsection{Multi-particle closed systems}

We start by assuming a one-dimensional ideal gas of $N$ non-identical 
classical particles with masses $m_i$, with $i=1,\ldots,N$, and total 
energy $E$. If particle $i$ has a momentum $m_iv_i$, newly we define a 
kinetic energy $K_i$ given by Eq. (\ref{eq-p_i}), where $p_i$ is the square 
root of $K_i$. If the total energy is defined as $E \equiv R^2$, we have 
\begin{equation}
p_1^2+p_2^2+\cdots +p_{N-1}^2+p_N^2 = R^2.
\label{eq-E1}
\end{equation} 
We see that the isolated system evolves on the surface of an $N$-sphere. 
The formula for the surface area $S_N(R)$
of an $N$-sphere of radius $R$ is
\begin{equation}
S_N(R) = {2\pi^{N\over 2}\over \Gamma({N\over 2})}R^{N-1},
\label{eq-S_n4}
\end{equation}
where $\Gamma(\cdot)$ is the gamma function. If the ergodic hypothesis is 
assumed, that is, each point on the $N$-sphere is equiprobable, 
then the probability $f(p_i)dp_i$ of finding 
the particle $i$ with coordinate $p_i$ (energy $p_i^2$) is proportional to the 
surface area formed by all the points on the $N$-sphere having the $i$th-coordinate 
equal to $p_i$. 
Our objective is to show that $f(p_i)$ is the Maxwellian 
distribution, with the normalization condition (\ref{eq-p_n2}).

If the $i$th particle has coordinate $p_i$, the $(N-1)$ remaining particles 
share the energy $R^2-p_i^2$ on the $(N-1)$-sphere
\begin{equation}
p_1^2+p_2^2 \cdots +p_{i-1}^2 + p_{i+1}^2 \cdots +p_N^2= R^2-p_i^2,
\label{eq-E11}
\end{equation} 
whose surface area is $S_{N-1}(\sqrt{R^2-p_i^2})$. 
If we define the coordinate $\theta$ as satisfying
\begin{equation}
R^2\cos^2\theta = R^2-p_i^2,
\label{eq-theta12}
\end{equation}
then
\begin{equation}
Rd\theta = {dp_i \over (1-{p_i^2\over R^2})^{1/2}}.
\label{eq-diftheta}
\end{equation}
It can be easily proved that
\begin{equation}
S_N(R) = \!\int_{-\pi/2}^{\pi/2}\!S_{N-1}(R\cos\theta)R d\theta.
\label{eq-theta14}
\end{equation}
Hence, the surface area of the $N$-sphere for which the $i$th coordinate is
between $p_i$ and $p_i+dp_i$ is $S_{N-1}(R\cos\theta)Rd\theta$.
We rewrite the surface area 
as a function of $p_i$, 
normalize it to satisfy Eq.~(\ref{eq-p_n2}), and obtain
\begin{equation}
f(p_i) = {1\over S_N(R)}
{S_{N-1}(\sqrt{R^2-p_i^2})\over (1-{p_i^2\over R^2})^{1/2}},
\label{eq-f_n3}
\end{equation}
whose final form, after some calculation is
\begin{equation}
f(p_i) = C_N R^{-1}\Big(1-{p_i^2\over R^2} \Big)^{N-3\over 2},
\label{eq-mm}
\end{equation}
with
\begin{equation}
C_N = {1\over\sqrt{\pi}}{\Gamma({N\over 2})\over \Gamma({N-1\over 2})}.
\label{eq-cn}
\end{equation}
For $N\gg 1$, Stirling's approximation can be applied to 
Eq.~(\ref{eq-cn}), leading to
\begin{equation}
\lim_{N\gg 1} C_N \simeq {1\over\sqrt{\pi}}\sqrt{N\over 2}.
\label{eq-cc}
\end{equation}
If we call $\epsilon$ the mean energy per particle, 
$E=R^2=N\epsilon$, then in the limit of large $N$ 
we have
\begin{equation}
\lim_{N\gg 1}\left(1-{p_i^2\over R^2}\right)^{N-3\over 2}
\simeq e^{-{p_i^2/2\epsilon}}.
\label{eq-ee}
\end{equation}
As in the former section, the Boltzmann factor $e^{-{p_i^2/2\epsilon}}$ is found 
when $N\gg 1$ but, even for small $N$, it can be a good approximation 
for particles with low energies.
After substituting Eqs.~(\ref{eq-cc})--(\ref{eq-ee})
into Eq.~(\ref{eq-mm}), we obtain the Maxwellian distribution (\ref{eq-gauss})
in the asymptotic regime $N\rightarrow\infty$ 
(which also implies $E\rightarrow\infty$).

Depending on the physical situation the mean energy per particle $\epsilon$
takes different expressions. For an isolated one-dimensional gas
we can calculate the dependence of $\epsilon$ on the temperature, which
in the microcanonical ensemble is defined by differentiating the entropy
with respect to the energy. The entropy can be written as $S=-kN\!\int_{-\infty} 
^{\infty} f(p)\ln f(p)\,dp$, where $f(p)$ is given by Eq.~(\ref{eq-gauss})
and $k$ is the Boltzmann constant. 
If we recall that $\epsilon=E/N$, we obtain
\begin{equation}
S(E)= {1\over 2}kN\ln\left({E\over N} \right) + {1\over 2}kN(\ln(2\pi)-1).
\end{equation}
The calculation of the temperature $T$ gives
\begin{equation}
T^{-1}= \left({\partial S\over \partial E} \right)_N = {kN\over 2E} = {k\over 2\epsilon}.
\end{equation}
Thus $\epsilon=kT/2$, consistent with the equipartition theorem. 
If $p^2$ is replaced by ${1\over 2}mv^2$, the Maxwellian 
distribution is a function of particle velocity, as it is usually given
in the literature:
\begin{equation}
g(v)dv = \sqrt{m\over 2\pi kT}\,e^{-{mv^2/2kT}}dv.
\end{equation}

\section{Conclusion}
\label{S4} \vspace{-4pt}

We have shown that the Boltzmann factor describes 
the general statistical behavior of each small part 
of a multi-component system whose components or parts are given
by a set of random variables that satisfy an additive constraint,
in the form of a conservation law (closed systems) or 
in the form of an upper limit (open systems).
The derivation of this factor for open systems in a general context
has been presented in \cite{lopez2009}.

Let us remark that these calculations do not need the knowledge of the exact or microscopic 
randomization mechanisms of the multi-agent system in order to reach the equiprobability.
In some cases, it can be reached by random forces \cite{yako2001}, in other cases 
by chaotic \cite{pellicer2009} or deterministic \cite{gonzalez2009} causes.
Evidently, the proof that these mechanisms generate equiprobability is not a trivial task
and it remains as a typical challenge in this kind of problems.

In summary, this work has presented a straightforward geometrical argument 
that recalls us the equivalence between canonical and microcanonical ensembles in 
the thermodynamic limit in the particular context of physical sciences.
For the general context of homogeneous multi-agent systems,
we conclude by highlighting the statistical equivalence 
of the volume-based and surface-based calculations in this type of systems.

\vspace{10pt} \noindent
{\bf Acknowledgements:} This research was supported by the
spanish Grant with Ref. FIS2009-13364-C02-C01.

\end{document}